\documentclass[prl,aps,twocolumn,nofootinbib,10pt, showpacs]{revtex4-1}
\usepackage{amssymb}
\usepackage{appendix}
\usepackage{times}
\usepackage{graphicx}
\usepackage{subeqnarray}

\usepackage{color}

\begin{document}

\title {Two-dimensional topological nodal line semimetal in layered $X_2Y$ ($X$ = Ca, Sr, and Ba; $Y$ = As, Sb, and Bi)}

\author{Chengwang Niu$^1$}
\email{c.niu@fz-juelich.de}
\author{Patrick M. Buhl$^1$}
\author{Gustav Bihlmayer$^1$}
\author{Daniel Wortmann$^1$} 
\author{Ying Dai$^2$}
\author{Stefan Bl\"{u}gel$^1$}
\author{Yuriy Mokrousov$^1$}
\affiliation
{$^1$Peter Gr\"{u}nberg Institut and Institute for Advanced Simulation, Forschungszentrum J\"{u}lich and JARA, 52425 J\"{u}lich, Germany
\\$^2$School of Physics, State Key Laboratory of Crystal Materials, Shandong University, 250100 Jinan, People's Republic of China}	

\begin{abstract}
In topological semimetals the Dirac points can form zero-dimensional and one-dimensional manifolds, as predicted for Dirac/Weyl semimetals and topological nodal line semimetals, respectively. Here, based on first-principles calculations, we predict a topological nodal line semimetal phase in the two-dimensional compounds $X_2Y$ ($X$=Ca, Sr, and Ba; $Y$=As, Sb, and Bi) in the absence of spin-orbit coupling (SOC) with a band inversion at the M point. The mirror symmetry as well as the electrostatic interaction, that can be engineered via strain, are responsible for the nontrivial phase. In addition, we demonstrate that the exotic edge states can be also obtained without and with SOC although a tiny gap appears at the nodal line for the bulk states when SOC is included.  
\end{abstract}

\maketitle
\date{\today}

Topological semimetals (TSMs), as new topological states of matter in addition to topological insulators (TIs)~\cite{Hasan,Qi1} and topological crystalline insulators (TCIs)~\cite{Ando}, have drawn enormous interest in recent years. Two prominent examples of TSMs are Dirac semimetals~\cite{Wang1,Wang2,Liuz,Neupane,Borisenko,Liuca,Jeon} and Weyl semimetals~\cite{Wan,Burkov,Xug,Huang,Weng1,Xu1,Lv,Yang1}. Their bulk states are gapless, conduction and valence bands touch only at discrete points in the Brillouin zone. The Dirac points in Dirac semimetals are composed of two Weyl points of opposite topological charge, which can be split by breaking either time-reversal symmetry or inversion symmetry, realizing so-called Weyl semimetals~\cite{Wan,Burkov,Xug,Huang,Weng1,Xu1,Lv,Yang1}. Like Dirac points in surface/edge states of TIs and TCIs, the set of Dirac (Weyl) points of Dirac (Weyl) semimetals is zero-dimensional. Recent work indicates that a set of Dirac points can also form an one-dimensional~\cite{Schnyder,Heikkila} ring in a class of TSMs called topological nodal line semimetals (TNLSs)~\cite{Weng2,xie,Kim,Yu2,Mullen,ChenY,Bian,Zeng,LiR,Bian2}, in which the conduction and valence bands touch along distinct ``nodal" lines. The existence of three-dimensional (3D) TNLS states has been confirmed for PbTaSe$_2$~\cite{Bian2}, ZrSiS~\cite{Schoop}, and PtSn$_4$~\cite{YWu}. 

To date, TIs and TCIs are realized in both 3D and two-dimensional (2D) systems~\cite{Hasan,Qi1,Ando,Wang1}. 2D systems bear great potential for the investigation of exotic phenomena not available in 3D, such as the quantum spin or quantum anomalous Hall effects~\cite{Yu,Fang}. Recently, based on symmetry arguments it was shown that the 2D Dirac semimetals can also been symmetry-protected~\cite{Young}. In this work, based on first-principles calculations, we report material realizations of TNLSs in a family of 2D materials, Ca$_2Y$, Sr$_2Y$, and Ba$_2Y$ ($Y$=As, Sb, and Bi) triple layers, where the TNLS state is protected by the mirror symmetry $z\rightarrow-z$. The computed non-zero topological invariant and the emergent edge states demonstrate the TNLS phase clearly. 

The density functional calculations are performed using the generalized gradient approximation (GGA) of Perdew-Burke-Ernzerhof (PBE)~\cite{Perdew} for the exchange correlation potential as implemented in the FLEUR code~\cite{fleur} as well as in the Vienna ab-initio simulation package (VASP)~\cite{Kresse,Kresse1} for relaxation. For each configuration, the atomic positions and lattice parameters were optimized until the forces are smaller than 0.01 eV/\AA ~with an energy cutoff of 500 eV.  A 20 \AA ~thick vacuum layer is used to avoid interactions between nearest slabs in VASP while the calculations are carried out with the film version of the FLEUR code~\cite{Krakauer}. From this the maximally localized Wannier functions (MLWFs) are constructed using the wannier90 code~\cite{Mostofi,Freimuth}. 

 \begin{figure}[!b]
\includegraphics{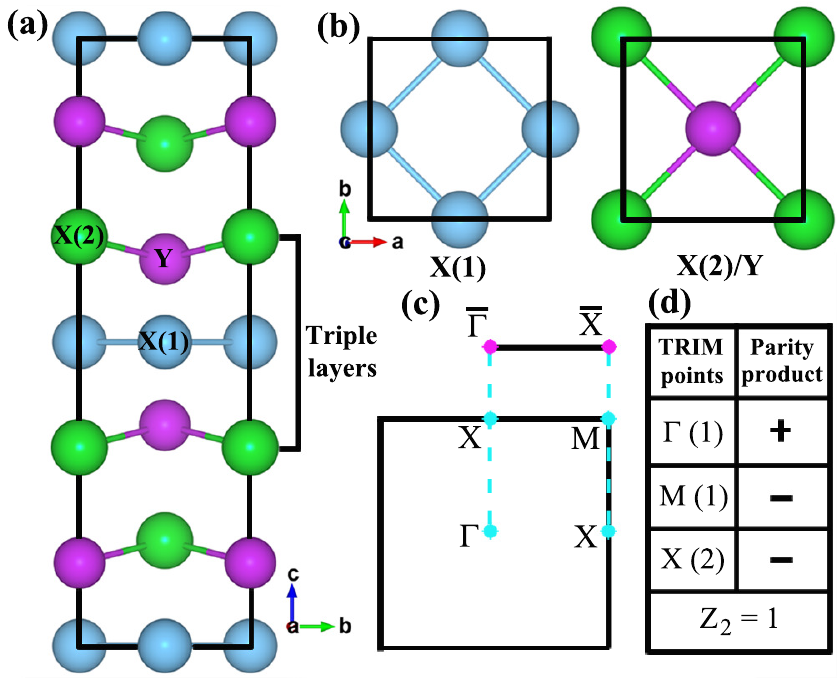}
\caption{
(a) Bulk $X_2Y$ ($X$=Ca, Sr, and Ba; $Y$=As, Sb, and Bi) crystallize in a tetragonal structure, space group $I4/mmm$, with the alternate stacking of $X(1)$ and $X(2)/Y$ layers. (b) Top view of $X(1)$~(left) and $X(2)/Y$~(right) layers. (c) Brillouin zone of the 2D triple layer and the projected 1D Brillouin zone. (d) Parity product of occupied states at TRIM points and the $\mathbb{Z}_2$ index for Ca$_2$As.} 
\label{structure}
\end{figure}

 \begin{figure*}
\includegraphics{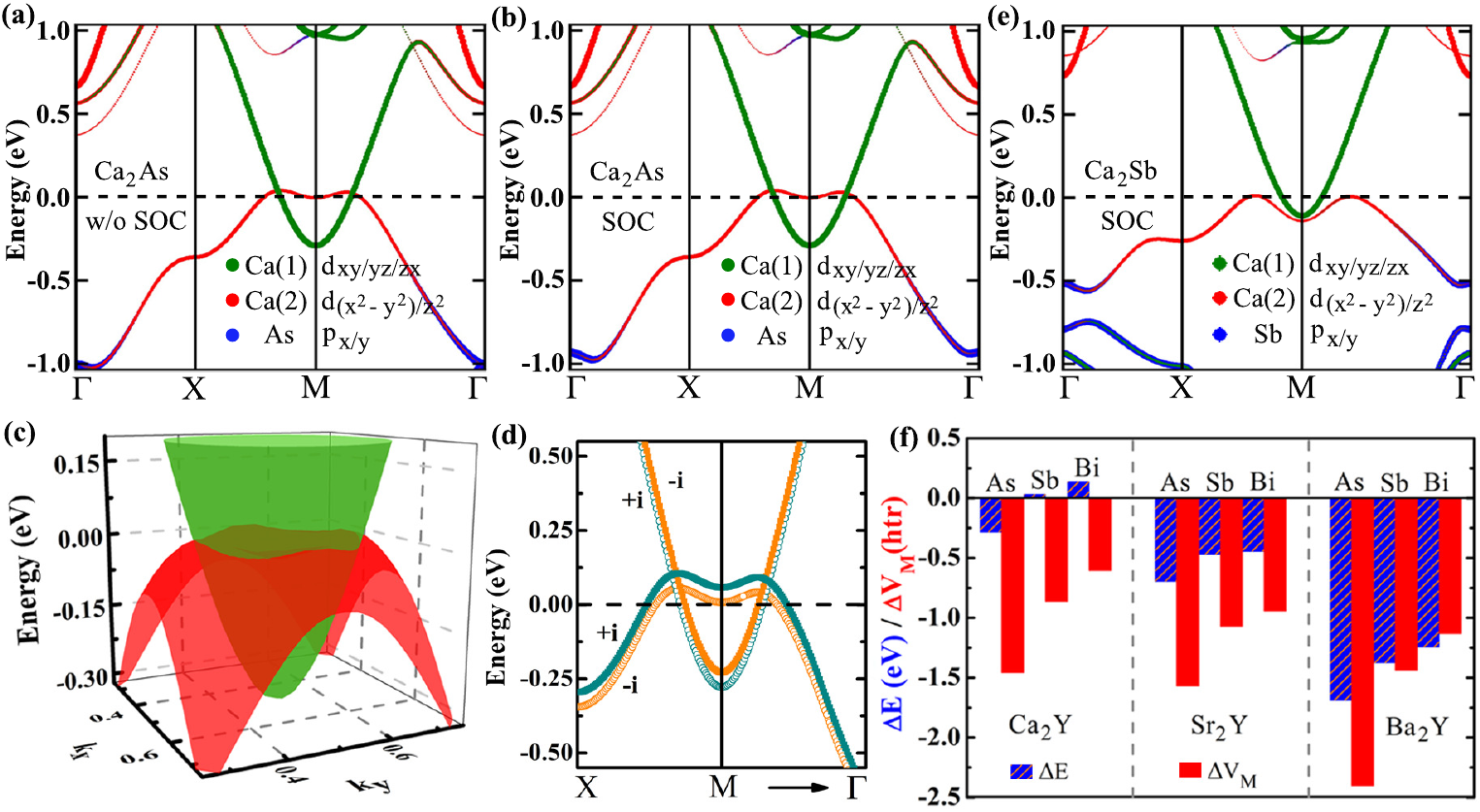}
\caption{
Orbitally-resolved band structures for Ca$_2$As triple layers (a) without and (b) with SOC, weighted with the Ca(1)-$d_{xy/yz/zx}$, Ca(2)-$d_{(x^2-y^2)/z^2}$, and As-p$_{xy}$ characters. The band dispersions and the orbital contributions are almost the same in both cases, but a tiny gap is opened when SOC is included. (c) Band crossings of the bands near the Fermi level form a nodal line, which is protected by the mirror symmetry $z\rightarrow-z$, and (d) the bands can be assigned a mirror eigenvalue $\pm i$ around the M-point. Two pairs of crossing bands are denoted by solid squares and empty circles respectively, and the solid squares have been shifted up by 50 meV for visibility. (e) Orbitally-resolved band structure of Ca$_2$Sb trilayer with SOC. A global gap appears around the M-point. (f) Inversion energy $\Delta E$ ($\Delta E = E_{X(1)}-E_{X(2)}$) at the M-point, and relative Madelung potential $\Delta V_M$ ($\Delta V_M=V_{X(1)}-V_{X(2)}$) for $X_2Y$ trilayers.}
\label{band}
\end{figure*}

Bulk $X_2Y$ compounds ($X$ = Ca, Sr, and Ba; $Y$ = As, Sb, and Bi) have tetragonal structure with space group $I4/mmm$ and have already been synthezized experimentally~\cite{Martinez,Pearson}.  As shown in Fig.~\ref{structure}(a), the crystal structure can be visualised as an alternating stacking of $X(1)$ and $X(2)/Y$ layers along the $c$-axis. $X(1)$ atoms are coplanar corresponding to the $D_{2h}$-site symmetry, while the $X(2)$ and $Y$ atoms exhibit buckling with one of the two neighbouring $X(2)/Y$ layers translated by a vector ($a$/2+$b$/2). One $X(1)$ layer and two $X(2)/Y$ layers constitute the simplest thin film in an ideal stoichiometry $X_2Y$. For films with nonideal stoichiometry, additional bands are introduced around the Fermi level and band dispersions become quite complex (Figure S1 in Supplemental Material~\cite{sup}). With ideal stoichiometry, $X_2Y$, total energy investigations show that the favourable triple layer configuration consists of one $X(1)$ layer enclosed by $X(2)/Y$ layers, invariant under $z\rightarrow-z$ reflection  [see Fig.~\ref{structure}(a)]. For example, the $z$-reflection symmetric Ca$_2$As trilayer is energetically more favorable by as much as 1.4 eV as compared to the structure with the Ca(1) layer on top of two Ca(2)As layers. 

To show the existence of DNLs in $X_2Y$ triple layers, we take Ca$_2$As as example and analyze the projected band structure for this system without and with SOC in Figs.~\ref{band}(a) and (b), respectively. In the absence of SOC, Ca(2)-$d_{(x^2-y^2)/z^2}$ and Ca(1)-$d_{xy/yz/zx}$ orbitals overlap around the M-point and the orbital characters exchange after passing through the band crossing points, indicating a band inversion with the Ca(1)-$d_{xy/yz/zx}$ band being lower by 0.29 eV than the Ca(2)-$d_{(x^2-y^2)/z^2}$ band. The gapless DNLs are realized as in the case of 3D Ca$_3$PdN~\cite{Kim,Yu2} when SOC is ignored. The three-dimensional vizualization of the nodal ring around the M-point is presented in Fig.~\ref{band}(c). Taking SOC into account, as shown in Fig.~\ref{band}(b), band dispersion remains almost the same around the Fermi level, but a tiny splitting of the band-crossing of about 2 meV is introduced (Figure S2 in Supplemental Material~\cite{sup}). 

The observed band inversion agrees well with the analysis of topological $\mathbb{Z}_2$ invariant~\cite{Kim}. For a Ca$_2$As trilayer with space inversion symmetry, the parity eigenvalues at the four time-reversal invariant momenta ($\Gamma$, 2X, M) give the products $\sigma_{\Gamma} = 1$ and $\sigma_X = \sigma_M= -1$, as shown in Fig.~\ref{structure}(d). This yields $\mathbb{Z}_2= 1$ thus confirming the topologically non-trivial feature and an odd number of DNLs in the 2D Brillouin zone. The DNL is protected by the mirror symmetry $z\rightarrow-z$, and this mirror symmetry implies that, in the mirror plane, the two crossing bands can be classified in terms of the mirror eigenvalues $\pm i$. Figure~\ref{band}(d) shows this clearly. Due to the presence of both time-reversal and inversion symmetries, the bands are degenerate when SOC is taken into account. The two crossing bands have opposite mirror eigenvalues in the vicinity of original crossing points although a tiny SOC-induced gap appears. The mirror symmetry protection is further explicitly confirmed by the DNL that survives (without SOC) under a perturbation that keeps the mirror symmetry $z\rightarrow-z$ while it breaks the inversion and all kinds of rotational symmetries (Figure S3 in Supplemental Material~\cite{sup}). In fact, similarly to topological crystalline insulators, a band gap in the system can be induced by breaking the crystal mirror symmetry. We demonstrate this by slightly moving Ca(2) and As atoms (by 0.46 \AA) at one side of the layer along the $z$-direction, and observing a gap opening at the original gapless crossing points (Figure S4 in Supplemental Material~\cite{sup}). When considering the films consisting of an integer number of Ca$_2$As trilayers (Figure S5 in Supplemental Material~\cite{sup}), we observed a non-trivial gap opening for an even number (when the mirror symmetry is broken), while the gapless DNLs survive for the odd number of the trilayers (when the mirror symmetry is preserved). 

Similar nodal lines appear around the Fermi energy in Sb$_2Y$ and Ba$_2Y$, but a global energy gap opens in Ca$_2$Sb and Ca$_2$Bi (Figure S6 in Supplemental Material~\cite{sup}). In Fig.~\ref{band}(e), which displays the projected band structure of Ca$_2$Sb including SOC, we observe that the lowest unoccupied bands are dominated by Ca(1)-$d_{xy/yz/zx}$ states, while the highest occupied bands at the M-point are dominated by the states of Ca(2)-$d_{(x^2-y^2)/z^2}$ character. There is no band inversion around the M-point in Ca$_2$Sb and Ca$_2$Bi, although they share the same lattice structure as well as symmetries with Ca$_2$As, Sb$_2Y$, and Ba$_2Y$. This situation is analogous to typical TCIs such as SnTe family~\cite{Hsieh}, where SnTe is topologically non-trivial, while PbTe is a trivial insulator. 

\begin{figure}
\includegraphics{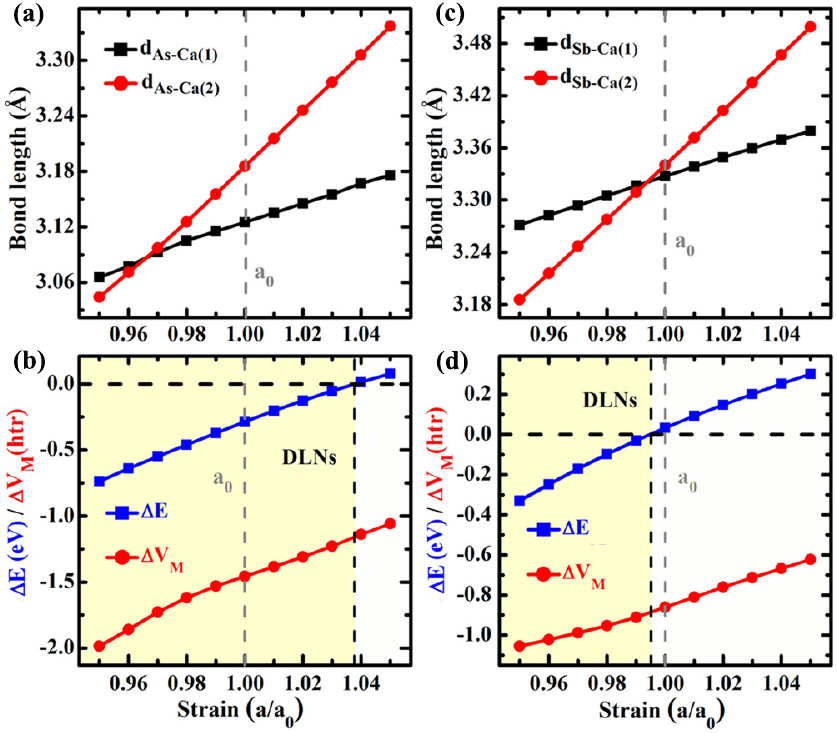}
\caption{Realization of a band inversion via strain for (a,b) Ca$_2$As and (c, d) Ca$_2$Sb. (a,c) Bond lengths, (b, d) inversion energy $\Delta E$, and relative Madelung potential $\Delta V_M$ can be modulated effectively by hydrostatic strain with an accompanying phase transition between nontrivial and trivial semimetals. The negative $\Delta E$ represent an inverted band and an emergence of a DNL at the M-point.}
\label{strain}
\end{figure}

To investigate the origin of the DNLs in $X_2Y$ further, we now focus on the role that the eletrostatic interaction plays for their formation. Namely, we find that  the energetic arrangement of the bands in this family relates directly to  the electronegativity difference of the constituents $\Delta\chi$($X_2Y$), $\Delta\chi$($X_2Y$) = $\chi(X)$ - $\chi(Y)$. Following the sequence $|$$\Delta\chi$($X_2$As)$|$ $>$ $|$$\Delta\chi$($X_2$Sb)$|$ $>$ $|$$\Delta\chi$($X_2$Bi)$|$, the inversion energy at the M-point, $\Delta E= E_{X(1)}-E_{X(2)}$, increases and signals an inverted band arrangement at M when it is negative (see Fig.~\ref{band}(f) and Table SI in Supplemental Material~\cite{sup}). This trend explains the absence of the band inversions in Ca$_2$Sb and Ca$_2$Bi despite a 
stronger SOC of Bi and Sb atoms as compared to that of As.  We further compute the Madelung potentials on $X(1)$ ($V_{X(1)}$) and $X(2)$ ($V_{X(2)}$) sites, whose contributions form the DNLs. We define the relative Madelung potential $\Delta V_M$ as $\Delta V_M=V_{X(1)}-V_{X(2)}$. A clear correlation of the trend in $\Delta E$ and $\Delta V_M$, presented in Fig.~\ref{band}(f), indicates that the electrostatic interaction between $X$ and $Y$ atoms plays an important role for the band inversion as well as the formation of DNLs in the considered family of compounds, besides the mirror-symmetry.

We further explore the stability of the band inversion with respect to strain, which is a very effective way to modulate the topological properties. Strain-induced phase transitions between normal insulators, TIs, and TCIs have been clearly confirmed~\cite{Chadov,YangK,NiuC}. The magnitude of the in-plane strain is described by $a/a_0$ ratio, where $a_0$ and $a$ denote the lattice parameters of the unstrained and strained systems, respectively. Internal atomic positions under homogeneous in-plane strain are fully relaxed. The corresponding As-Ca bond length $d_{\rm {As-Ca}}$, inversion energy $\Delta E$, and relative Madelung potential $\Delta V_M$ of Ca$_2$As versus the hydrostatic strain are presented in Figs.~\ref{strain}(a) and (b). As can be seen in this figure, 
all computed quantitites are quite sensitive to strain.  As the lattice parameters are increased, the As-Ca(1) and As-Ca(2) distances increase as well, resulting in a decreasing interaction between As and Ca states. However, as shown in Fig.~\ref{strain}(a), $d_{\rm {As-Ca(2)}}$ increases much faster than $d_{\rm {As-Ca(1)}}$, and the variation of the corresponding Ca-As interaction is different for Ca(1) and Ca(2). This is clearly reflected in  $\Delta V_M$, see Fig.~\ref{strain}(b). In case of an increasing bond length both $V_{X(1)}$ and $V_{X(2)}$ decrease, but since $V_{X(2)}$ changes more rapidly, $\Delta V_M$ increases in accordance with $\Delta E$ so that the DNLs disappear under tensile strain stronger than 3.8\%. A similar trend is observed for Ca$_2$Sb [see Figs.~\ref{strain}(c) and (d)], for which the DNLs can been obtained by a compressive strain of about $-$0.5\%. In the latter case a strain-induced band inversion occurs at the M-point due to the increase in the electrostatic interaction. 

\begin{figure}
\includegraphics{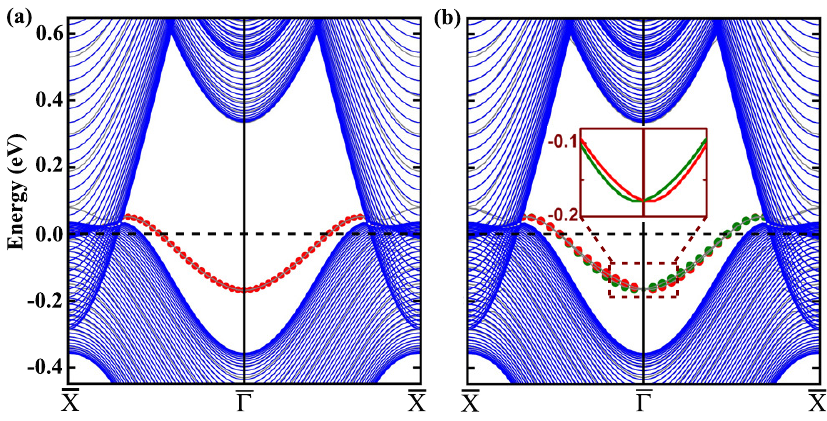}
\caption{ 
Band structures of a Ca$_2$As nanoribbon with Ca(1)-As termination (gray lines) and the projected bulk band structure (blue lines) (a) without and (b) with SOC. Inset in panel (b) shows the zoom-in view at the $\bar\Gamma$ point. The circle size denotes the weight of the outmost four layers at one edge and the color indicates the spin direction.}
\label{edge}
\end{figure}

At this point, we are ready to demonstrate the emergence of characteristic edge states, that are another manifestation of the non-trivial topology of considered systems. Taking Ca$_2$As as an example, a symmetric nanoribbon with approximately 75 \AA  ~width with Ca(1)-As termination is constructed, and the dangling bonds of edge atoms are saturated by Li. The calculated band structures of nanoribbon as well as projected bulk band structures are presented in Fig.~\ref{edge}. In the case without SOC, as shown in Fig.~\ref{edge}(a), one can easily see the emergence of topological edge states. Taking into account SOC, Fig.~\ref{edge}(b) shows that a SOC-induced spin splitting apears for the topological edge states, which cross each other at $\Gamma$ point and carry opposite spin polarizations. This is quite similar to surface/edge states of TI.

\begin{figure}
\includegraphics{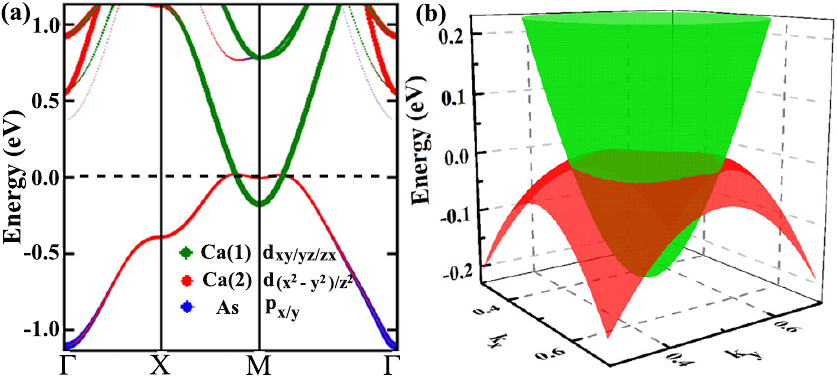}
\caption{ 
(a) Orbitally resolved band structure of Ca$_2$As sandwiched between two KCl layers, weighted with the Ca(1)-$d_{xy/yz/zx}$, Ca(2)-$d_{(x^2-y^2)/z^2}$, and As-p$_{xy}$ characters. The band-crossing remains even with the substrate present. (b) 3D representation of the band structure in (a) which shows the DNL around the M-point.}
\label{kcl}
\end{figure}

Finally, keeping the possibilty for experimental observation in mind, we explore the possible realization of DNLs on a lattice-matching substrate and propose a quantum well structure which keeps the mirror symmetry in the system. For Ca$_2$As trilayer, we select KCl films as cladding layers with a lattice mismatch of 3.9\%. Figure~\ref{kcl}(a) shows the orbitally resolved band structure of Ca$_2$As sandwiched between two KCl layers with optimized lattice constants. As can be seen, the energy bands with the Ca(2)-$d_{(x^2-y^2)/z^2}$ orbital character overlap around the M point with the Ca(1)-$d_{xy/yz/zx}$ states, in good qualitative agreement with free-standing Ca$_2$As. The crossing line of the conduction and valence bands in the 2D Brillouin zone is also explicitly shown in Fig.~\ref{kcl}(b) in the vicinity of the M-point. We also examine the feasibility of quantum well structure by calculating the interface formation energy, defined as $2E_{\rm f} = E_{\rm QW} - E_{ \rm Ca_2As} - 2E_{\rm KCl}$ , where $E_{\rm QW}$ is the total energy of the quantum well structure, while $E_{ \rm Ca_2As}$ and $E_{\rm KCl}$ are the energies of thin films. The calculated interface formation energy for Ca$_2$As sandwiched between two KCl layers is $E_{\rm f} = -0.66$~eV i.e. of a size comparable with the interface formation energy of two Ca$_2$As trilayers (-0.73 eV). Thus a layered growth and the formation of a quantum well structure seems realistic. In addition, the DNL around the M-point remains intact even though the KCl lattice parameters correspond to a 3.9\% tensile strain (Figure S7 in Supplemental Material~\cite{sup}). Remarkably, according to our calculations, we can state that the DNLs are obtained for a wide range of cladding layer thickness (Figure S8 in Supplemental Material~\cite{sup}).

In summary, we theoretically predict that $X_2Y$ ($X$=Ca, Sr, and Ba; $Y$=As, Sb, and Bi) triple layers realize a family of 2D topological nodal line semimetals protected by crystalline symmetry. A nonzero topological index and the emergence of the edge states demonstrate the nontrivial phase clearly. In addition, we reveal the possibility of the 2D topological node-line semimetal states on a lattice-matching substrate. As various techniques have been developed to grow various 2D materials with complex structure, we expect that $X_2Y$ trilayers and proposed QWs can be synthesized in experiments via MBE and realize the 2D nontrivial phase presented above.

Note~added. Recently, we became aware of three independent work on the prediction of Hg$_3$As$_2$~\cite{Lu}, PdS~\cite{Jin}, and Be$_2$C~\cite{Yangb} as 2D TNLSs without SOC.

We are grateful for helpful discussions with J.-W. Rhim. This work was supported by the Priority Program 1666 of the German Research Foundation (DFG), the Virtual Institute for Topological Insulators (VITI), the National Basic Research Program of China (973 program, 2013CB632401), and the Taishan Scholar Program of Shandong Province. We acknowledge computing time on the supercomputers JUQUEEN and JURECA at J\"{u}lich Supercomputing Centre and JARA-HPC of RWTH Aachen University.

\end{document}